\documentclass[a4paper,12pt]{article}
\usepackage{amsmath,amssymb}
\usepackage{latexsym}
\usepackage{graphicx}
\usepackage{authblk}

\newtheorem{theorem}{Theorem}
\newtheorem{corollary}[theorem]{Corollary}
\newtheorem{proposition}[theorem]{Proposition}
\newtheorem{observation}[theorem]{Observation}

\newtheorem{lemma}[theorem]{Lemma}

\begin{document}

\title{Non-crossing geometric spanning trees
with bounded degree and monochromatic leaves
on bicolored point sets}

\author[1]{Mikio Kano\thanks{This work was started when the author visited Universidad de Alcal\'a in May and June of 2018, and this work was supported by Universidad de Alcal\'a (Giner de los R\'{\i}os grant) and JSPS KAKENHI Grant Number 16K05248.}}

\author[2]{Kenta Noguchi\thanks{This work was supported by JSPS KAKENHI grant Number 17K14239.}}

\author[3]{David Orden\thanks{This work was supported by project MTM2017-83750-P of the Spanish Ministry of Science (AEI/FEDER, UE), as well as by H2020-MSCA-RISE project 734922 - CONNECT.}}

\affil[1]{Ibaraki University, Hitachi, Ibaraki, Japan \\
e-mail: mikio.kano.math@vc.ibaraki.ac.jp}

\affil[2]{Tokyo University of Science, Noda, Chiba, Japan \\
e-mail: noguchi@rs.tus.ac.jp}

\affil[3]{Universidad de Alcal\'a, Alcal\'a de Henares, Madrid, Spain \\
e-mail: david.orden@uah.es}

\date{}

\maketitle

\begin{abstract}
Let $R$ and $B$ be a set of red points and a set of blue points in the plane, respectively, such that $R\cup B$ is in general position, and let $f:R \to \{2,3,4, \ldots \}$ be a function. We show that if $2\le |B|\le \sum_{x\in R}(f(x)-2) + 2$,
then there exists a non-crossing geometric spanning tree $T$ on $R\cup B$ such that
$2\le \deg_T(x)\le f(x)$ for every $x\in R$ and the set of leaves of $T$ is $B$, where every edge of $T$ is a straight-line segment.
\end{abstract}

\section{Introduction and related work}
Let $R$ be a set of red points and $B$ be a set of blue points in the plane. We always assume that $R$ and $B$ are disjoint and $R\cup B$ is in general position (i.e, no three points of $R\cup B$ are collinear). Several works~\cite{Abellanas-1997,Kano-Suzuki-2013} have considered problems on non-crossing geometric spanning trees and geometric graphs (where edges are straight-line segments) on $R\cup B$. See also the survey~\cite{Kaneko-Kano-2003}.

\subsection{Contributions of this work}

For a tree $T$ and a vertex $v$ of $T$, let us denote by $\deg_T(v)$ the degree of $v$ in~$T$. A vertex of $T$ with degree one is called a {\em leaf} of $T$, and the set of leaves of $T$ will be denoted by $Leaf(T)$. Further, let us denote by $|T|$ the order of a tree $T$ (i.e., its number of vertices) and by $|X|$ the cardinality of a set~$X$.

\begin{figure}[htbp]
 \begin{center}
 \includegraphics*{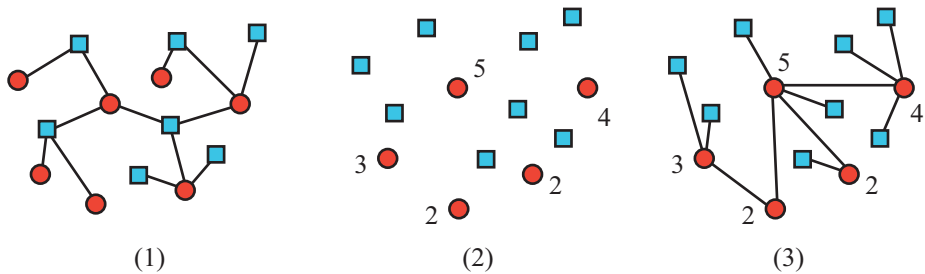}
 \end{center}
 \caption{(1) A non-crossing geometric alternating spanning tree on 7 red points and 7 blue points with maximum degree 3;  (2) A red point set $R$ with labels $f(x)$ and a blue point set $B$;  (3) A non-crossing geometric spanning tree $T$ on $R\cup B$ such that $Leaf(T)=B$ and  $\deg_T(x)=f(x)$ for every $x\in R$.}
 \label{fig-1}
 \end{figure}

In this paper, given $R$ and $B$ in the plane as above, we aim for non-crossing geometric spanning trees $T$ on $R\cup B$ such that $Leaf(T)=B$. We prove the following theorem:

\begin{theorem}  Assume that $R$ and $B$ are given in the plane and a function
$f:R \to \{ 2,3,4, \ldots\}$ is given. If $2\le |B| \le \sum_{x\in R}(f(x)-2)+2$, then there exists a non-crossing geometric spanning tree $T$ on $R\cup B$ such that $Leaf(T)=B$ and  $2\le \deg_T(x)\le f(x)$ for every $x\in R$. Moreover, if  $|B|= \sum_{x\in R}(f(x)-2)+2$, then $T$ satisfies that $\deg_T(x)=f(x)$ for every $x\in R$ (see (2) and (3) of Figure~\ref{fig-1}).
\label{th-4}
\end{theorem}

By setting $f(x)=k$ for every $x\in R$ in the above theorem,
we obtain the following corollary:

\begin{corollary}  Let $k\ge 2$ be an integer. Assume that $R$ and $B$ are given in the plane. If $2\le |B|\le (k-2)|R|+2$, then there exists a non-crossing geometric spanning tree $T$ on $R\cup B$ such that $Leaf(T)=B$ and the maximum degree of $T$ is at most $k$.
Moreover, if  $|B|= (k-2)|R|+2$, then $T$ satisfies that $\deg_T(x)=k$ for every $x\in R$.
\label{cor-5}
\end{corollary}

\begin{observation}
For $k=2$, the above corollary says that if $|B|=2$ and $|R|\ge 1$, then there exists a non-crossing geometric path which passes through all the points of $R\cup B$ and whose endvertices are the two blue points.
\end{observation}

\subsection{Related work}

Ikebe et al. \cite{Ikebe-1994} proved that, given one red point $R=\{r\}$ and a set $B$ of blue points in the plane, any rooted tree $T$ with root $w$ of order $|B|+1$ can be straight-line embedded on $\{r\}\cup B$ in such a way that $w$ is mapped to $r$ and no crossings arise. Kaneko and Kano~\cite{Kaneko-Kano-1999} proved that, given two red points $R=\{r_1,r_2\}$ and a set $B$ of blue points in the plane, together with two rooted trees $T_1$ with root $w_1$ and $T_2$ with root $w_2$, if $|T_1|+|T_2|=|B|+2$, then $T_1\cup T_2$ can be straight-line embedded on $\{r_1,r_2\}\cup B$, without crossings, in such a way that $w_1$ and $w_2$ are mapped to $r_1$ and $r_2$, respectively.

Kaneko \cite{Kaneko-1998} considered sets
$R$ and $B$ in the plane with $|R|=|B|$ and proved that, then, there exists a non-crossing geometric spanning tree $T$ on $R\cup B$ such that every edge of $T$ joins a red point to a blue point and the maximum degree of $T$ is at most~3 (see (1) of Figure~\ref{fig-1}). Finally, Biniaz et al.~\cite{Biniaz-2018} considered sets $R$ and $B$ in the plane with $|B|\le |R|$ and proved that, then, there exists a non-crossing geometric spanning tree $T$ on $R\cup B$ such that every edge of $T$ joins a red point to a blue point and the maximum degree of $T$ is at most $\max\{3, \lceil (|R|-1)/|B|\rceil +1\}$.


\section{Proof of Theorem~\ref{th-4}}

In this section we prove Theorem~\ref{th-4}.
We first state the following proposition, which is a special case of Theorem~\ref{th-4}.

\begin{proposition}
Assume that $R$ and $B$ are given in the plane and a function $f:R \to \{ 2,3,4, \ldots\}$ is given. If $|B|=\sum_{x\in R}(f(x)-2)+2$, then there exists a non-crossing geometric spanning tree $T$ on $R\cup B$ such that $Leaf(T)=B$ and  $\deg_T(x)= f(x)$  for every $x\in R$ (see (2) and (3) of Figure~\ref{fig-1}).
\label{prop-6}
\end{proposition}

In order to prove Proposition~\ref{prop-6} we will use the following lemma:

\begin{lemma}[Theorem~3.6 of \cite{Chartrand}, and Exercises~2.1.12 of \cite{West}] Let $n\ge 2$ be an integer, and let $d_1, d_2, \ldots, d_n$ be positive integers. If $d_1+d_2+ \cdots +d_n=2n-2$, then there exists a tree $T$ with vertex set $\{v_1,v_2, \ldots, v_n\}$ that satisfies $\deg_T(v_i)=d_i$ for every $1\le i \le n$.
\label{lem-7}
\end{lemma}

\medskip \noindent
 {\em Proof of Proposition~\ref{prop-6}}. We fist show that there exists a geometric spanning tree $Q$ on $R\cup B$ that might have crossings but satisfies
\begin{equation}
 \deg_Q(x)=f(x) \quad \mbox{for all $x\in R$,~~and~~ } \deg_Q(y)=1 \quad \mbox{for all $y\in B$.}
\label{eq-10}
\end{equation}
It follows from the condition of Proposition~\ref{prop-6}  that
\begin{align*}
 & \sum_{x\in R}f(x)+\sum_{y\in B} 1 = \sum_{x\in R}(f(x)-2)+2|R|+|B| \\
 = &  |B|-2+2|R|+|B|=2|R\cup B|-2.
\end{align*}
Hence by Lemma~\ref{lem-7}, there exists a geometric spanning tree $Q$ that satisfies
the condition (\ref{eq-10}) but might have some crossings.
Among all geometric spanning trees $Q$ satisfying (\ref{eq-10}), choose a geometric spanning tree $T$ on $R\cup B$ such that the sum $\sum_{xy\in E(T)} |xy|$ is minimum, where $|xy|$ denotes the length of the straight-line edge $x$ to $y$. We shall show that $T$ has no crossings.

\begin{figure}[htbp]
 \begin{center}
 \includegraphics*{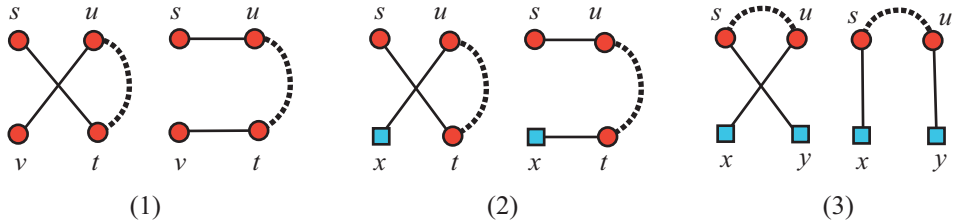}
 \end{center}
 \caption{ (1) Two intersecting edges $st$ and $uv$ and two new edges $su$ and $vt$, which satisfy $|st|+|uv|> |su|+|vt|$; (2) Two crossing edges $st$ and $ux$ and two new edges $su$ and $tx$, where $x$ is a blue point and a leaf of $T$; (3)  Two crossing edges $sy$ and $ux$ and two new edges $sx$ and $uy$, where $x$ and $y$ are blue points and  leaves of $T$.}
 \label{fig-2}
 \end{figure}

The following three possible types of crossings could arise.  First, that two edges $st$ and $uv $ of $T$ intersect, where $s,t,u,v$ are red points  (see (1) of Figure~\ref{fig-2}). Since $T-st-uv$ consists of three components,
by symmetry, we may assume that $u$ and $t$ are contained in the same component of
$T-st-uv$, that is, $u$ and $t$ are connected by a path in $T-st-uv$. Then $T-st-uv+su+vt$ is another geometric spanning tree on $R\cup B$ satisfying the degree condition (\ref{eq-10})  and its total  sum of edge-lengths is smaller than that of $T$. This contradicts the choice of $T$. Hence this case does not occur.

Second, that two edges $st$ and $ux$ of $T$ intersect, where $s,t,u$ are red points and $x$ is a blue point (see (2) of Figure~\ref{fig-2}). Since $T-st-ux$ consists of three components and $\{x\}$ forms one component, $u$ and $t$ are connected by a path in $T-st-ux$ or
$u$ and $s$ are connected by a path in $T-st-ux$. By symmetry, we may assume that $u$ and $t$ are connected by a path in $T-st-ux$. Then $T-st-ux+su+tx$ is another geometric spanning tree on $R\cup B$  satisfying the degree condition (\ref{eq-10}) and its total sum of edge lengths is smaller than that of $T$. This is a contradiction.

Third, that two edges $sy$ and $ux$ of $T$ intersect, where $s,u$ are red points and $x,y$ are blue points (see (3) of Figure~\ref{fig-2}). Since $T-sy-ux$ consists of three components and $\{x\}$ and $\{y\}$ form two components,  $s$ and $u$ are connected by a path in $T-sy-ux$. Then $T-sy -ux +sx+uy$ is another geometric spanning tree on $R\cup B$  satisfying the degree condition (\ref{eq-10}) and its total  sum of edge-lengths is smaller than that of $T$. This is a contradiction.

Note that blue points being leaves implies that these three were the only possible cases for crossings and, therefore, $T$ has no crossings. Consequently, $T$ is the desired non-crossing geometric spanning tree on $R\cup B$, and Proposition~\ref{prop-6}  is proved. ~~$\Box$

\medskip
We next prove Theorem~\ref{th-4} by making use of Proposition~\ref{prop-6}.

\medskip \noindent
 {\em Proof of Theorem~\ref{th-4}}. We may assume that $2\le |B| < \sum_{x\in R}(f(x)-2)+2$ since if  $|B|=\sum_{x\in R}(f(x)-2)+2$, then the theorem holds
by Proposition~\ref{prop-6}.
It is easy to see that there exists a mapping $f': R \to \{2,3,4, \ldots\}$ that satisfies
$f'(x)\le f(x)$ for all $x\in R$ and $|B|=\sum_{x\in R}(f'(x)-2)+2$.
By Proposition~\ref{prop-6}, there exists a non-crossing geometric spanning tree $T$ such that $Leaf(T)=B$ and $\deg_T(x)=f'(x)$ for all $x\in R$. Hence, $T$ is the desired geometric spanning tree on $R \cup B$.  Consequently Theorem~\ref{th-4} is proved. ~~ $\Box$

\end{document}